\documentclass[12pt]{article}
\usepackage{amssymb}
\usepackage{amsmath}
\usepackage{epsfig}
\usepackage{graphicx}
\usepackage{wrapfig}

\begin{document}

\title{ The study of a new gerrymandering methodology
\footnote{Supported by our deep interests in mathematical
modeling.}} \vspace{3mm}

\author{{ Pan Kai, Tan Yue, and Jiang Sheng }\\
{\small  Department of Modern Physics, University of Science and Technology}\\
{\small  of China (USTC), Hefei, Anhui 230026, P.R.China} }

\date{}
\maketitle \vskip 12mm

\begin{abstract}
This paper is to obtain a simple dividing-diagram of the
congressional districts, where the only limit is that each district
should contain the same population if possibly. In order to solve
this problem, we introduce three different standards of the "simple"
shape. The first standard is that the final shape of the
congressional districts should be of a simplest figure and we apply
a modified "shortest split line algorithm" where the factor of the
same population is considered only. The second standard is that the
gerrymandering should ensure the integrity of the current
administrative area as the convenience for management. Thus we
combine the factor of the administrative area with the first
standard, and generate an improved model resulting in the new
diagram in which the perimeters of the districts are along the
boundaries of some current counties. Moreover, the gerrymandering
should consider the geographic features.The third standard is
introduced to describe this situation. Finally, it can be proved
that the difference between the supporting ratio of a certain party
in each district and the average supporting ratio of that particular
party in the whole state obeys the Chi-square distribution
approximately. Consequently, we can obtain an archetypal formula to
check whether the gerrymandering we propose is fair.
\end{abstract}

\newpage
\tableofcontents

\newpage

\section{Introduction}

\begin{figure}[hbt]
\begin{center}
  \includegraphics[width= 12 cm]{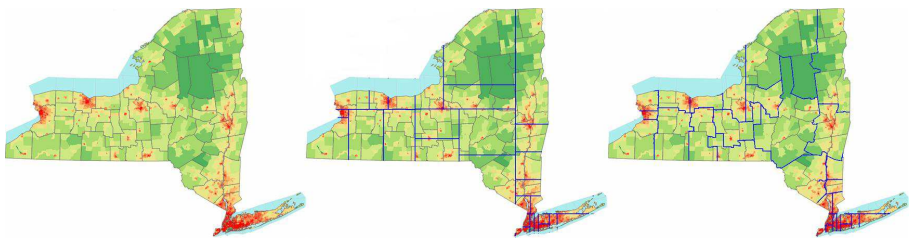}\\
  \caption{\label{result} (a)The map of population density$^{[1]}$. (b)The shape of districts produced by simple model. (c)The shape of districts produced by improved model}
\end{center}
\end{figure}

\noindent To ensure the fairness of the election, the foremost task
is to obtain a scientific arrangement of the boundaries of the
congressional districts. Although the states constitution provides
the number of representatives each state may have, it articulates
nothing about how the district shall be determined geographically.

\vspace{1em}This oversight has nowadays led to an unnatural district
shape which includes many long and narrow areas. Taking the state of
New York as an example, on the map of current congressional
districts$^{[2]}$, we can see that there are some unnatural-shaped
districts, such as districts 20$^{th}$ and 22$^{nd}$ (shown in
Figure \ref{conclusion}(a)). To create the ``simplest" shapes, where
the only limit is that each district should contain the same
population if possibly, we introduce three different standards of
the simplicity:

\begin{enumerate}
\item Each district is approximatively a rectangle.
\item Based on the above standard, the outlines are best along the boundaries of counties.
\item The modifications of geographical features and fairness are
combined in the final model.
\end{enumerate}

Therefore, we should take into accounts comprehensive factors
including the population, the boundaries of counties, geographic
features and fairness when gerrymandering.

\subsection{Issues in the Model}
\begin{itemize}
\item The first objective is to divide the state into some districts
following the rule that the population of each district must be
same.
\item As the second objective, we wish to make the boundaries of the
districts as simple as possible according to the 1$^{st}$ standard
we have proposed above.
\item The third objective is that we need further modifications
considering the standard 2 and standard 3.
\end{itemize}

\subsection{Previous Works}
There are many existing methods to realize the gerrymandering, with
the consideration of different aspects. For instance, the ``shortest
split line algorithm"$^{[3]}$ considers the simplicity of the shape
as key factors. Another method, which uses the statistical physics
approach, takes every county as a single element of a matrix, makes
an analogue to the q-state pots$^{[7]}$ model, and gets the optimal
solution keeping the integrality of a county. Other approaches such
as the ``fixed district algorithm"$^{[3]}$ and the ``changing the
voting system method"$^{[3]}$ mainly consider the unbiasedness of
the election as the crucial factor.

\vspace{1em}In our work, the foremost task is to make sure the
simplicity of the shapes of districts. Therefore, we modify the
``shortest split line algorithm" to establish our initial simple
model. The main modification is that we only use horizontal line and
vertical line instead of diagonals. In this way, we can get simple
rectangles other than irregular polygonal districts. While in the
section of Model Modification we develop a new method to ensure the
perimeter of the districts are along the boundaries of the current
counties.

\subsection{Our Approach}

\begin{itemize}
\item First, we obtain the population density
matrix from the density map. \item In order to obtain a simple
shape of districts, we only take into account the rule that every
district should have the same population to establish a simple
model. \item Then we optimize our model by adding the modification
of the present shapes of counties and gain a more nature looking
figure.\item Based on the results from the forgoing simulation, we
finally investigate into the factor of geography and fairness in
detail when gerrymandering.
\end{itemize}
\subsection{Our Result}
\vspace{1em}According to the steps mentioned above, Figure
\ref{result} shows our results when applying our model to the state
of New York.

\section{Basic Gerrymandering Model}
In order to establish our basic model, we first obtain a population
density matrix using the population map. We then establish a simple
model with the only constraint being that each district have the
same population. Finally, we optimize our model by adding additional
constraints such as preventing the division of original
administrative areas.

\subsection{Acquirement of the Population Density Matrix}
To calculate the population each district contains, we must first
extract the population density matrix for later computer programming
from the existing census data of some big cities and the macroscopic
population density map.

\begin{figure}[htb]
\begin{center}
  \includegraphics[width= 8 cm]{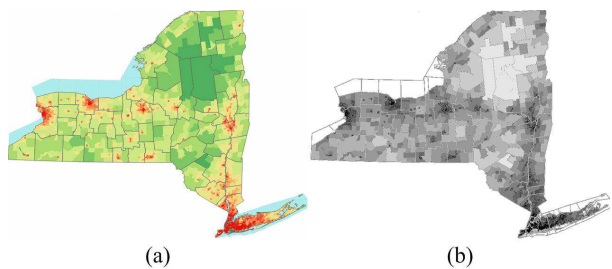}\\
  \caption{\label{grey} (a)The colored map of population density. (b)The grey figure of population density}
\end{center}
\end{figure}

\vspace{1em}For the original colored population map, we use Matlab
to identify the color of every pixel. Thus we can get the population
density corresponding to the particular pixel though which we obtain
the population density matrix. Moreover, this matrix can be shown as
a grey level figure(Figure \ref{grey}).

\subsection{A Simple Model for Redistricting}

In order to obtain simple shapes for congressional districts, we
modified the original ``shortest split line algorithm" and developed
a simple algorithm to divide districts. In our model:

\begin{itemize}
\item We focus on keeping an even distribution of population in each district, ignoring other
factors.
\item The algorithm takes as input only the population density matrix and ignores other factors such as party loyalties of the citizens, thus guaranteeing unbiased
results.
\item We use horizontal and vertical lines to separate the state result in fairly rectangle districts.
\end{itemize}

\begin{figure}[htb]
\begin{center}
  \includegraphics[width= 8 cm]{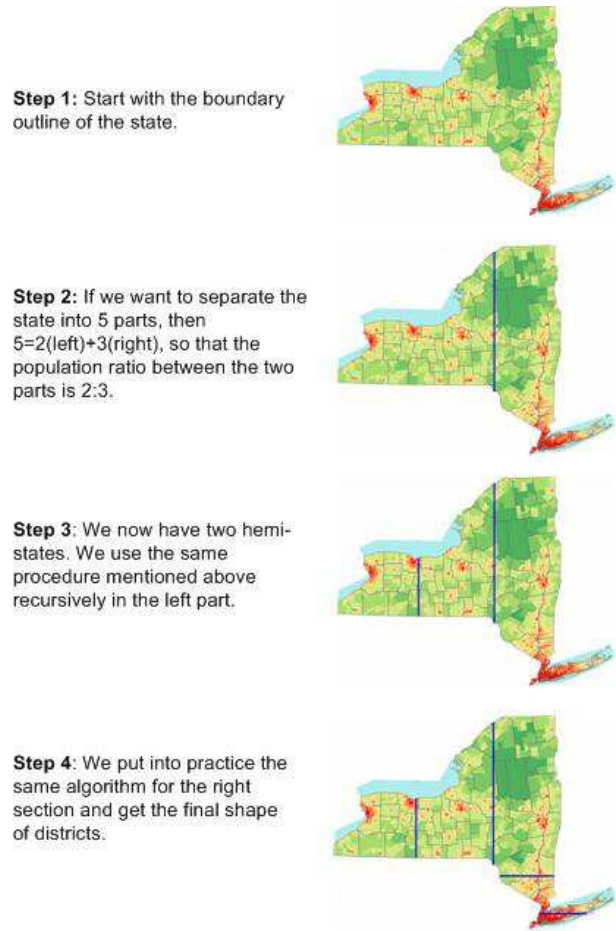}\\
  \caption{\label{pic2} The procedure of the algorithm of the simple model.}
\end{center}
\end{figure}

\vspace{1em}The procedure of the algorithm can be shown in Figure
\ref{pic2}.

\vspace{1em}This district-dividing algorithm has the advantage of
simplicity, clear unbiasedness, and it produces fairly nice-looking
rectangle districts.

\vspace{1em}{\bf The advantages of the simple model:}
\begin{itemize}
\item Simplicity
\item Clear unbiasedness
\item Fairly nice-looking rectangle districts.
\end{itemize}

{\bf The disadvantages:}
\begin{itemize}
\item Fails to take into consideration other factors such as geographic features
and integrality of the counties.
\end{itemize}

\subsection{Model Modification}

Despite of its advantage of simplicity, the original model has the
disadvantage of ignoring the shape of administrative area and
geographic features. As the figures show, the boundaries produced by
the original algorithm some times divide a county which is an
administrative area into different parts. This can prove
inconvenient and unnatural when carrying out the election
procedures. To dispel this disadvantage, we made the following
improvements to our original model:

\subsubsection{A Basic Assumption}
Before taking into consideration additional factors in dividing the
counties, we need to make the following assumption: we assume that
two districts share approximately the same population if the
difference in their numbers of voters is no more than $5 \%$.

\vspace{1em}In reality, it is impossible to divide districts into
absolutely equal numbers of voters due to the influence of
population flow and the fluctuation of birth and death rates. And
most of the constitutions also allow the the standard deviation of
number of voters to fall within $10-15 \%^{[7]}$.

\subsubsection{Modification Method}
After making up the designs of landform, we are faced with a problem
- how to adopt the borderline to avoid dividing up administrative
districts. To solve this problem, we have to adopt the borderlines
between the administrative districts. We follow the lines in bulge
and concave in some district. As we think the density in population
almost equals close to borderline between districts. Taking into
account the factors of population, landform and management, we take
this division as a simple meaning in management.

\subsubsection{Our Algorithm}

The basic consideration of this improved model is the same with the
above simple model. However, in that model, the dividing lines are
all straight and thus cannot keep a single county integrate. Now we
are trying to overcome this shortcoming and do our best to make the
dividing line coincide with the boundary of each administrative
area, although county dividing is unavoidable.

\vspace{1em}\noindent{\bf Step 1:}

Use the simple model to find the straight dividing line.
\begin{figure}[htb]
\begin{center}
  \includegraphics[width= 8 cm]{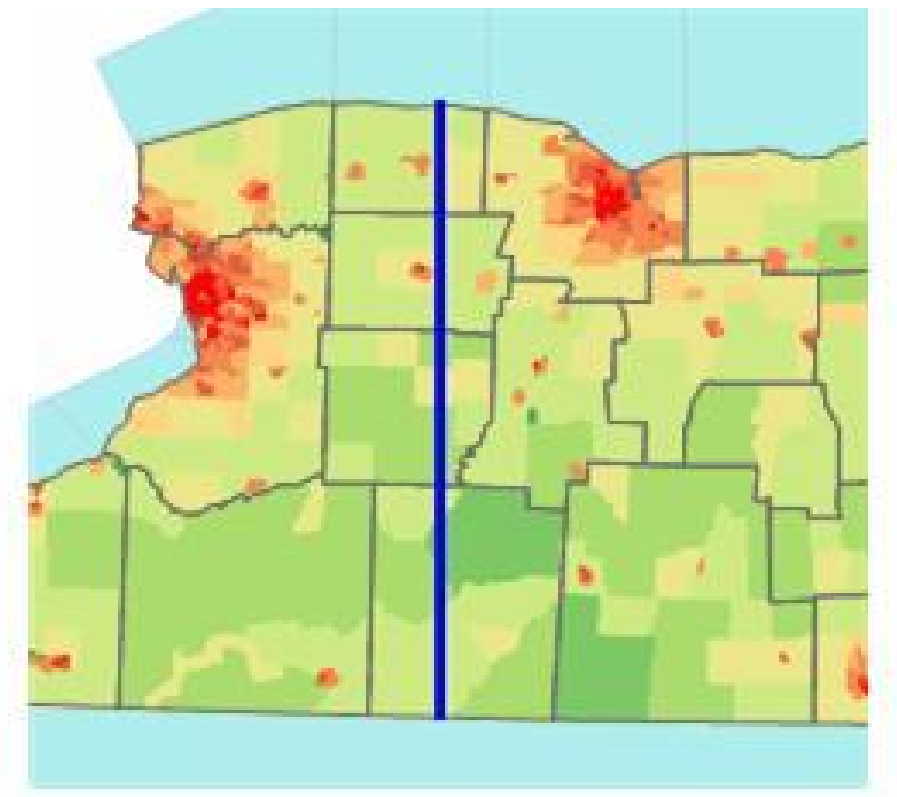}\\
  \caption{\label{step1} step 1}
\end{center}
\end{figure}

\newpage
\vspace{1em}\noindent{\bf Step 2:}

Find all the intersections between the straight line and the county
boundaries. Then in the program, use ``count" the integer to
represent how many crossover points there are on the straight line
and use array point[count] to record the location of all these
intersections.

\begin{figure}[htb]
\begin{center}
  \includegraphics[width= 8 cm]{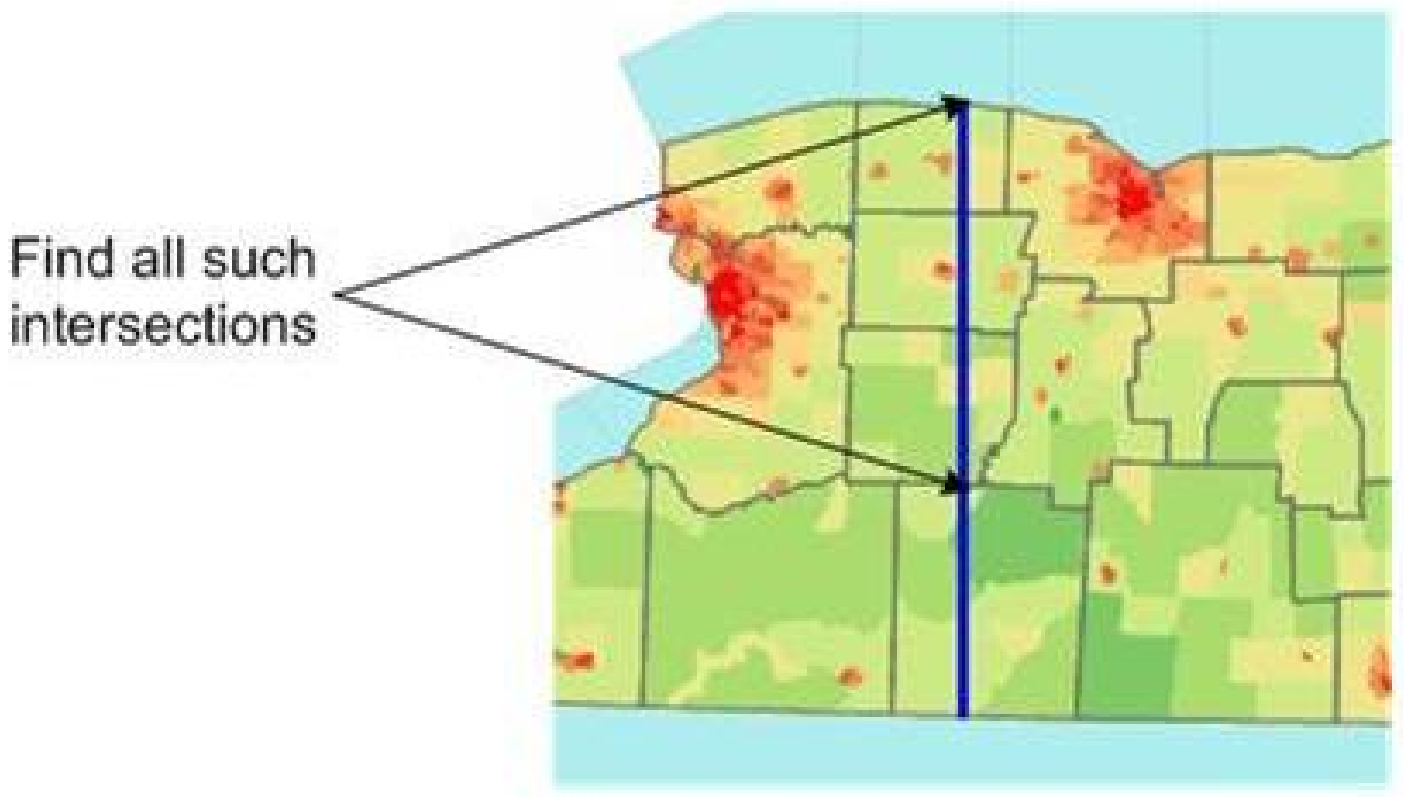}\\
  \caption{\label{step2} step 2}
\end{center}
\end{figure}

\vspace{1em}\noindent{\bf Step 3:}

There must be 2 or 3 direct paths along the straight line or the
boundary of the county the straight line goes through connecting the
two intersection points. We are trying to decide which of the two
paths along the boundary is more ``simple". The ``simple" path
should be the boundary which connects the adjoining points and more
close to the straight line.

\vspace{1em}First we find the point on the straight line whose
coordinate position is 2 or 3 units away from the intersection. We
separately search the points left and right to it (take vertical
line as an example, for horizontal lines the 2 directions would be
above and below). The direction we finally choose would be the one
on which we arrive at the boundary earlier. We record the direction
chosen in array path[count-1], if the direction is left or upward,
path[i]=1, if it is right or downward, path[i]=3.

\vspace{1em}This method may not be very strict, but it is easy to
realize and takes very little time for the computer to execute.

\begin{figure}[htb]
\begin{center}
  \includegraphics[width= 8 cm]{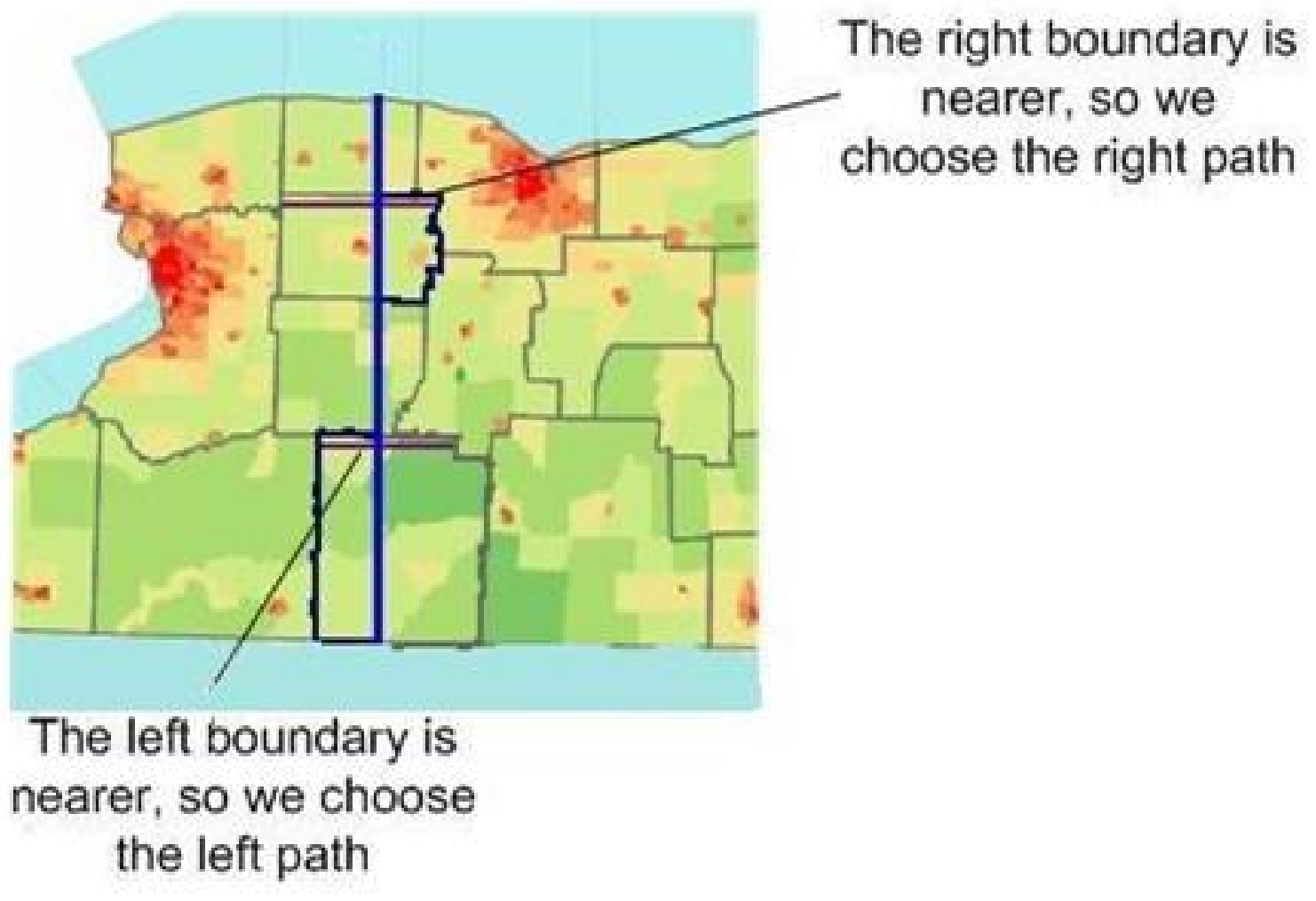}\\
  \caption{\label{step3} step 3}
\end{center}
\end{figure}

\vspace{1em}\noindent{\bf Step 4:}

Every straight line between two adjoining points and the boundary
line chosen in step 3 can confine a small area. We calculate the
population in each small area by sum up the elements of the
population density matrix including in the area. Pay attention that
if the area is to the left or below the straight line, prescribe
that s[i] is negative, otherwise, it is positive. We save the result
in array s[count-1].

\begin{figure}[htb]
\begin{center}
  \includegraphics[width= 8 cm]{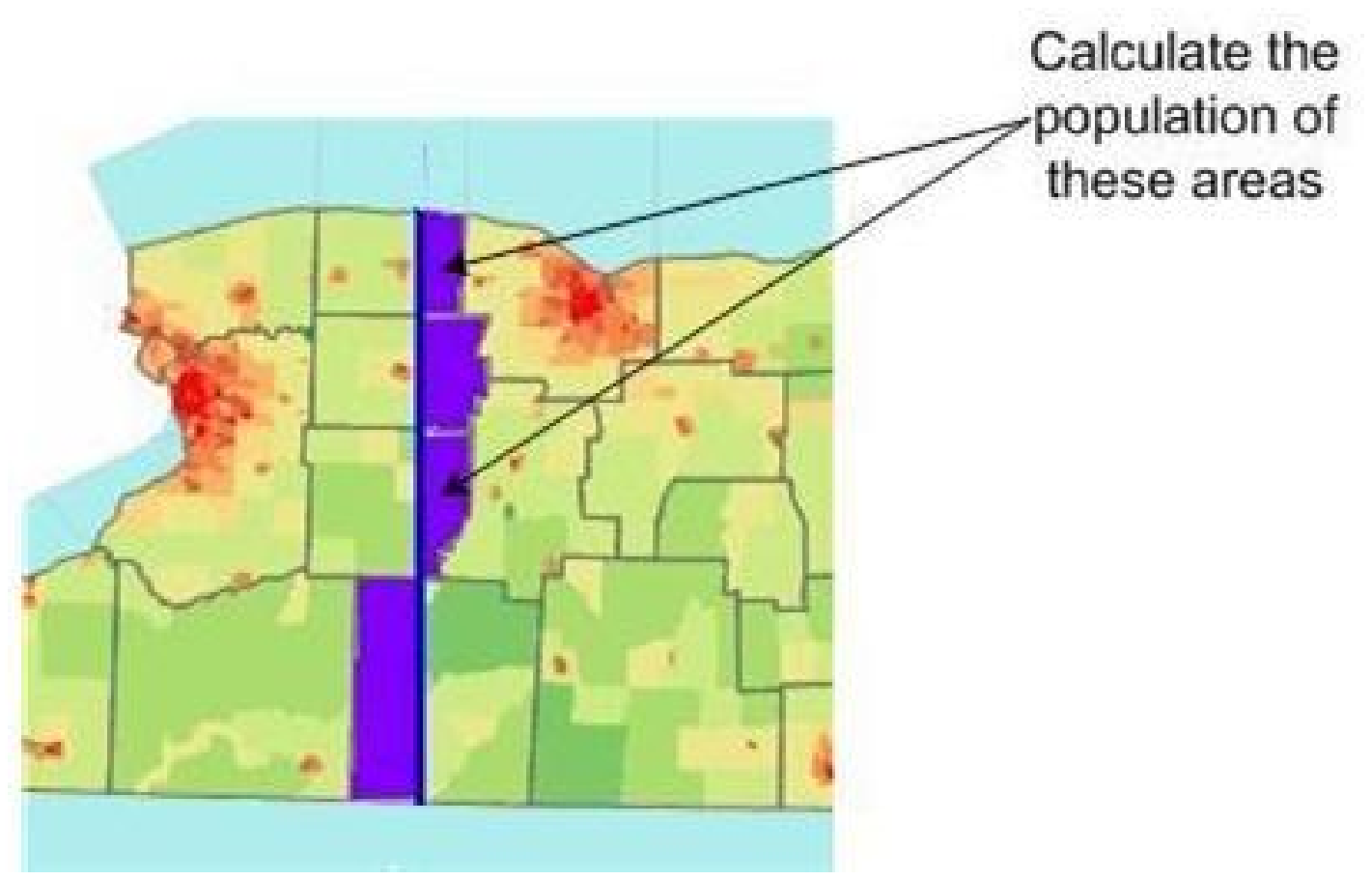}\\
  \caption{\label{step1} step 4}
\end{center}
\end{figure}

\vspace{1em}\noindent{\bf Step 5:}

In this step we will finally decide the dividing line - its shape
and location. We now know all the points on the dividing line and
want to decide the lines joining the points. We use array
path[count-1] to describe these lines. path[i] stands for the line
connecting the i$^{th}$ point and the (i+1)$^{th}$ point. path[i]
=1,2,3 stands for the $1^{st}$, $2^{nd}$, $3^{th}$ path count from
left to right or downwards to upwards.

\vspace{1em}The principles we use to decide which path to choose is
like this:

\vspace{1em}First, while we replace the initial straight line with
the boundary line, the population of different congressional
districts will vary to each other. So the differences must be
controlled within a certain range.

\vspace{1em}Second, we should use as little as possible straight
lines, because choosing a straight path means one county will be
divided into two.

\begin{figure}[htb]
\begin{center}
  \includegraphics[width= 8 cm]{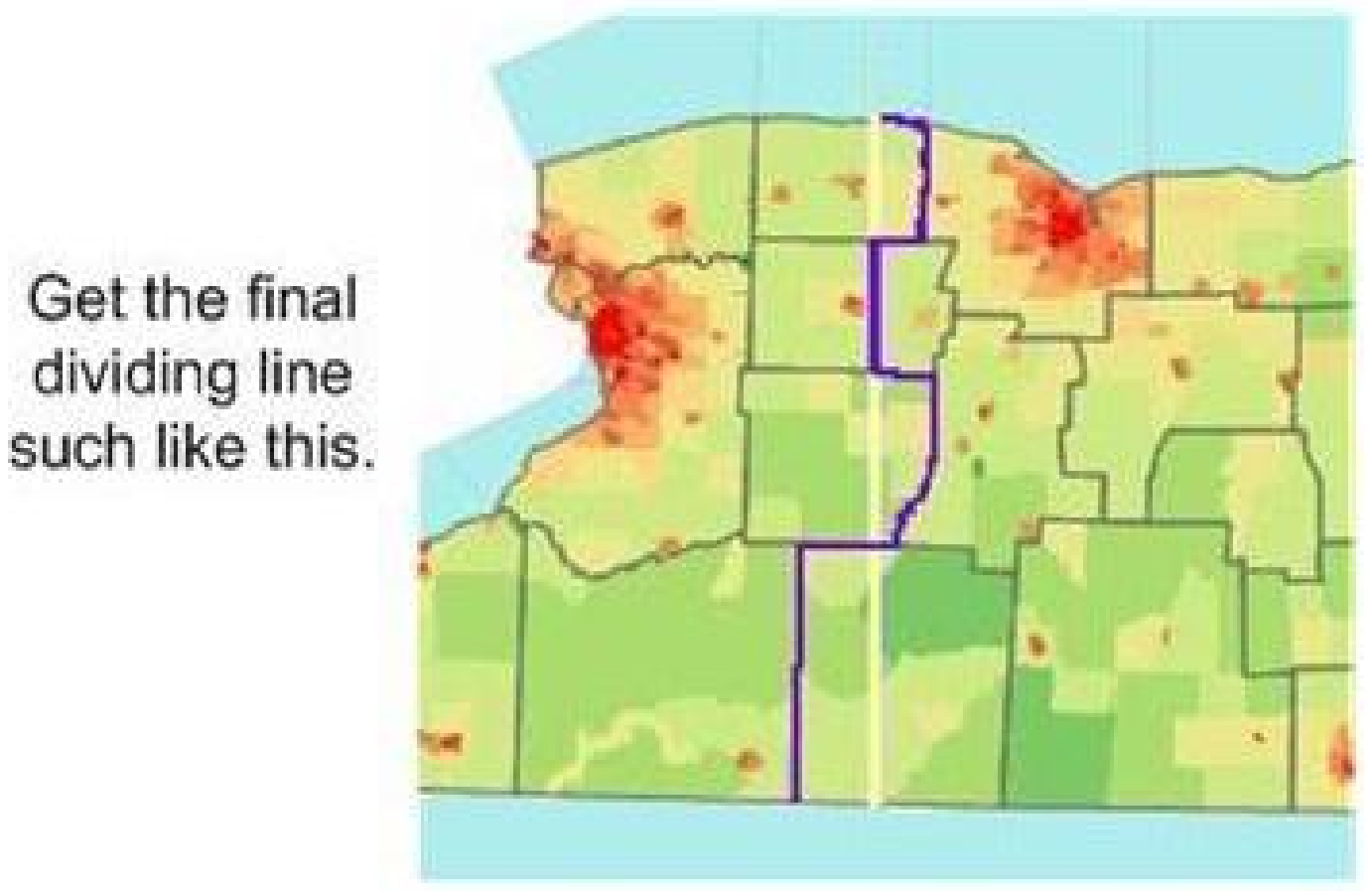}\\
  \caption{\label{step5} step 5}
\end{center}
\end{figure}

\vspace{1em}The method is as follows:

\vspace{1em}In step 4 we get the array s[count-1], we first sum up
all the elements in s[count-1] and get the result S. Judge whether
$S<delta$, here, delta can be the theoretical population of a
congressional district multiple with the allowed range of error (in
our program we choose the range as $5\%$), if so, we consider
current path is qualified and we get the final array path[count-1],
exit loop; if not, we find s[i] which is closest to S ( $|S - s[i]|$
is the minimum), and let path[i]=2 (change the path to the straight
line)£¬let $S=S-s[i]$, return to judge whether $S<delta$, ... ,and
loop like this.

\vspace{1em}After exiting the loop, print the location of the
initial straight line and its end points, point[count] and
path[count-1].

\vspace{1em}The final dividing line we get with this method may not
be the optimal line, but it is qualified. The method can avoid
considering all the possible combination of paths and thus can
operate very fast.

\vspace{1em}{\bf Step 6:} Do the recursion as the simple model and
get all the dividing line.

\section{Further Optimization of the Model}
Except shape and county integrity, other factors should also be
taken into account in actual gerrymandering. For example, how to
avoid biased ``gerrymandering¡±, how to modify our model in special
landform, as we are going to represent.

\subsection{The Modification of Avoiding ``Gerrymandering"}
\subsubsection{Brief Introduction}
There are two principal strategies behind gerrymandering$^{[3,8]}$:
maximizing the effective votes of supporters, and minimizing the
effective votes of opponents. One form of gerrymandering, packing is
to place as many voters of one type into a single district to reduce
their influence in other districts. A second form, cracking,
involves spreading out voters of a particular type among many
districts in order to reduce their representation by denying them a
sufficiently large voting block in any particular district. The
methods are typically combined, creating a few ``forfeit" seats for
packed voters of one type in order to secure even greater
representation for voters of another type.

\begin{figure}[htb]
\begin{center}
  \includegraphics[width= 8 cm]{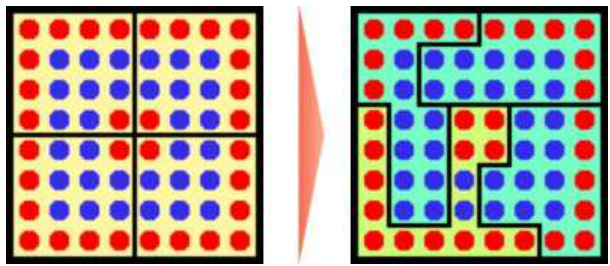}\\
  \caption{\label{pic8}$^{[3]}$ Redrawing the balanced electoral districts in this example
creates a guaranteed 3-to-1 advantage in representation for the blue
voters. Here, 14 red voters are packed into the yellow district and
the remaining 18 are cracked across the 3 blue districts.}
\end{center}
\end{figure}

\subsubsection{Resolution}
Because the simple algorithm we based on uses only the shape of the
state, the number m of districts wanted, and the population
distribution as inputs - and does not know the party loyalties of
the voters in any given region - the result cannot be
biased$^{[3,9]}$.

\vspace{1em}However we can not completely prevent the certain voter
biases from occurring due to the random redistricting process -
although the probability is small. As a result, we need make a
mathematical analysis to determine whether the gerrymandering our
model proposed is fair.

\begin{itemize}
\item {\bf Model assumption}:
\end{itemize}

Due to the the influence of population flow and the fluctuation of
birth, death and the supporting ratio of a political party. It is
reasonable to assume that the supporting ratio of a particular party
in each district is incorrelate of one another approximately. Next
we would analyze how to judge the fairness of the gerrymandering to
a specific party (Republican in our example) ; other parties and
situations can be judged via the same method.

\begin{itemize}
\item {\bf Mathematical Analysis}:
\end{itemize}

\begin{equation}
x_i^j=\left\{
\begin{array}{c c}
     1 & If\ the\ i^{th}\ voter\ in\ the\ j^{th}\ district\ support\ the\ Republican\\
     0 & Otherwise
\end{array}
\right.
\end{equation}

\vspace{1em}If $p_j$ represents the supporting ratio of the
Republican in $j^{th}$ district and this district has n voters, then
we would expect to get (3.2)

\begin{equation}
p_j=\frac{\sum_{i=1}^n x_i^j}{n}.
\end{equation}

\vspace{1em}Let p be the last-few-year average supporting ratio of
the Republican in the whole state. Thus we can expect that

\begin{equation}
p  \approx \frac{\sum_{i=1}^m p_j}{m}.
\end{equation}

\vspace{1em}Obviously, $x_i^j$ is distributed by B(1,p).

\begin{eqnarray}
X_j=\sqrt{\frac{n}{p(1-p)}}(p_j-p)&=&\sqrt{\frac{n}{p(1-p)}}(\frac{\sum_{i=1}^n x_i^j-np}{n})\nonumber\\
                              &=&\frac{\sum_{i=1}^n x_i^j-np}{\sqrt{np(1-p)}}\nonumber\\
                              &\sim&N(0,1)
\end{eqnarray}

\vspace{1em}According to the statistics theory, $X_j$ obeys N(0,1),
if n is large enough.

\vspace{1em}Then it is reasonable to obtain the following deduction:

\begin{equation}
Y=\sum_{j=1}^m X_j^2 \sim \chi_{m}^2,
\end{equation}

\vspace{1em}Here m indicates the number of districts in a certain
state. Thus, we can use Equ.(3.5) to generate a simple but effective
standard to judge the fairness of the original gerrymandering.

\vspace{1em}According to the statistics theory, if we get the
population density of the supporting ratio of the Republican, it is
easy to gain $Y$ using the data $p_j$ and p.

\vspace{1em}After that, the foremost task is to determine a
parameter $\alpha_{allow}$ scientifically. The parameter
$\alpha_{allow}$ functions as a threshold to judge whether the
current gerrymandering is bias or not. Then from the data table of
distribution $\chi^2$, we can get the value $\alpha$ where
$P(X>Y)=\alpha$. Here $X$ stands for the random variable distributed
by $\chi_m^2$.

\begin{figure}[htb]
\begin{center}
  \includegraphics[width= 12 cm]{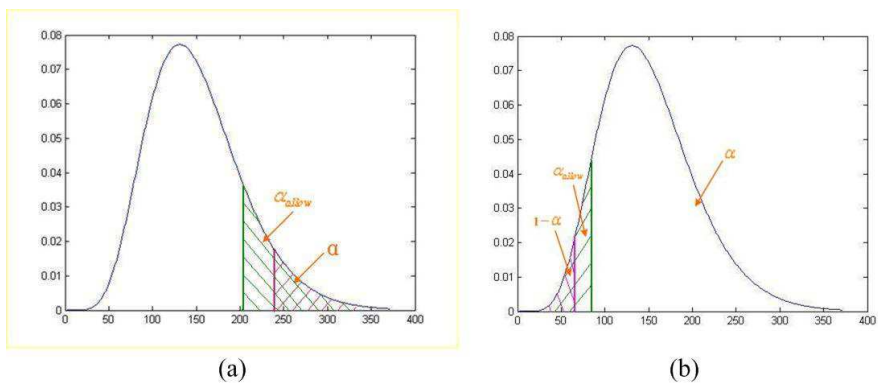}\\
  \caption{\label{pic9} (a)Packing strategy (b)Cracking strategy}
\end{center}
\end{figure}

\vspace{1em}On one hand, if there happens the ``packing" situation,
we would expect $\alpha$ to be extremely small. Therefore, if
$\alpha_{allow}>\alpha$, we can judge this gerrymandering as packing
situation leading to unfairness. On the other hand, if ``cracking"
situation happens, we can expect $1-\alpha$ to be extremely small.
We consider this gerrymandering biased as the same (shown in Figure
\ref{pic9}).

\vspace{1em}In conclusion, if the gerrymandering is fair, the value
of $\alpha$ produced by it would comply with following limits.

\begin{equation}
\begin{array}{cccl}
    \alpha &>& \alpha_{allow}& Not\ ``packing"\\
    1-\alpha &>& \alpha_{allow}& Not\ ``cracking".
\end{array}
\end{equation}

\begin{figure}[htb]
\begin{center}
  \includegraphics[width= 8 cm]{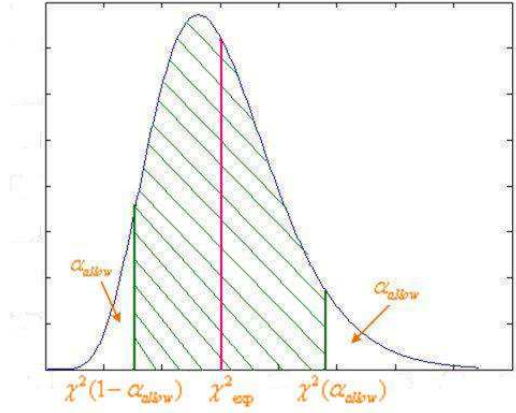}\\
  \caption{\label{x2newff} Permitted $Y$ Domain}
\end{center}
\end{figure}

\vspace{1em}We can briefly obtain the final limit on value $\alpha$
(shown in Figure \ref{x2newff}), that is
\begin{equation}
\alpha_{allow}<\alpha<1-\alpha_{allow}.
\end{equation}

\begin{itemize}

\item{\bf Conclusion:}
\end{itemize}
\vspace{1em}After generating the ``simple" congressional districts,
we can use Equ.(3.5) and (3.7) to judge whether it satisfies the
requirement of fairness, if we can collect the supporting ratio of
the Republican in each district, the last-few-year average
supporting ratio in the whole state and the parameter
$\alpha_{allow}$. If it is not a fair gerrymandering, we need to
reconsider the procedure of redistricting.

\subsection{The Modification of Geographic Features}

Up till now, many models that can realize gerrymandering have
existed. Thus comparison between all kinds of gerrymandering results
is necessary. We are going to further analysis the characteristic of
our two models while comparing the results we get with others.

\subsubsection{Brief Introduction}
Above-mentioned method has ignored landform factors such as main
rivers, mountains, lakes in process of choosing partition, thus it
has made district partition unnatural. Here we make some
improvements.

\vspace{1em}As to main rivers, mountains and so on, we hold the view
that along the river, population benefit in economy comparatively
resembles. On this basis, the areas on the river should be one
district.
\subsubsection{Resolution}
Our operations are as follows:
\begin{enumerate}
\item  A new district with a large population is based on
expanding along the river and mountains. Considering the advantages
of the long river or deep mountains, as well as convenience in
management, more districts will be formed if the population
increases.\item To the large lake, we expand the area into several
districts along the bank, for they are blessed with the same
interest in economy. \item To the area of mountains and plain, we
consider developing them separately as a result of different demands
in benefit. \item The same thinking is practical along the
coastline. We assemble people on the coast into several districts,
to separate the areas of coastline from inland to help the
development from the coastline to the inland. \item Considering the
approaching interest in economy, the area of islands and bylands is
being considered alone. A special zone will be built.
\end{enumerate}

\vspace{1em}With these factors of  landform being considered alone,
priority to every electoral district deducted, to the part remaining
again, we will carry out the previous simple operation according to
the simple model.

\section{Conclusion and Analysis}

\subsection{Comparison of Our Two Models}
We have fully established two models - the simple one and the
refined one, both of which are easy to realize. Each of these two
models has its own prominent advantages.

\vspace{1em}As to the simple model, first of all, the final shapes
of the congressional districts are all very simple, constructing
with horizontal and vertical lines only. Moreover, the simple model
guarantees that each district contains the same population if given
a precise population division.

\vspace{1em}While to the refined model, we can maintain the
integrity of every single county to a large extent, which ease the
management of actual voting. What¡¯s more, although the population
of districts varies from each other, the differences can be
controlled theoretically to any proposed precision.

\subsection{Comparison With Other Models}
There are many existing methods to realize the gerrymandering with
the consideration of different aspects. Here we compares our model
with the ``shortest split line algorithm" and the ``q-state pots
model".

\vspace{1em}From the theoretically analysis and the final result, we
can see that our method has its merits.

\vspace{1em}First, one important task for us is to make sure the
simplicity of the shapes of districts. Therefore, we modify the
``shortest split line algorithm" to establish our initial simple
model. The main modification is that we only use horizontal line and
vertical line instead of diagonals. In this way, we can get simple
rectangles other than irregular polygonal districts produced by
original ``shortest split line algorithm".

\vspace{1em}Secondly, we think the ``q-state potts model" is
excellent. The foremost characteristic of it is that this model
keeps the integrity of counties. However, the difference of district
populations can just be controlled within $15\%$, while this
difference in our model is no more than $5\%$.

\vspace{1em}In sum, the advantages of our model is that:
\begin{enumerate}
    \item Easy to carry out in the computer.
    \item Simple shapes of the districts.
    \item Combined with the factor of avoiding dividing up the current administrative
ares.
    \item The difference of district populations can be controlled in any precision.
\end{enumerate}

\subsection{Comparison With Current Districts}

\begin{figure}[htb]
\begin{center}
  \includegraphics[width= 10 cm]{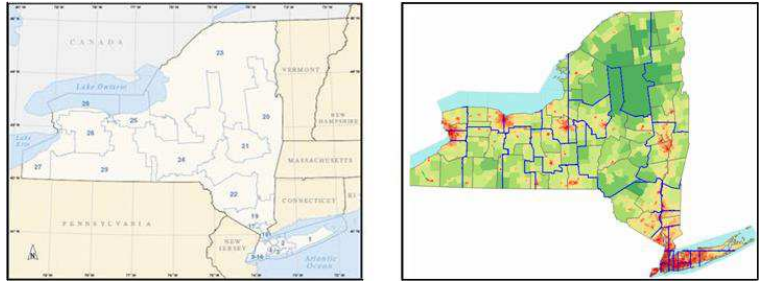}\\
  \caption{\label{conclusion} Comparison With Current Districts}
\end{center}
\end{figure}

\noindent From Figure \ref{conclusion}, we can clearly see the
difference between the current (the left one) districts and the
redistricting proposed by our model. Through careful comparison, we
can draw the conclusion that each has its advantages.

\vspace{1em}The current shape of districts own its great merit that
it take into the geographical consideration. For example, the
district $20^{th}$ is built along the Lake Ontario and the district
$22^{nd}$ mainly consists of plain. Both show the modification of
geographic features.

\vspace{1em}Although there is some merit in the current district,
the gerrymandering result produced by our model still show its
advantages.
\begin{enumerate}
    \item The shape of most districts are rectangular satisfying the
retirements of the ``simplicity".
    \item Most of the perimeters of the district are along the
boundaries of current counties resulting in the convenience of
management.
    \item Last but not least, the difference of district populations
is no more than $5\%$, and can be controlled in any precision.
\end{enumerate}

\vspace{1em}Finally, we have generated a simple but effective
criterion to judge the fairness of the gerrymandering via the
reasonably mathematical analysis which makes the model as a whole.

\section{Weakness and Further Development}

Although our two models are reasonable and easy to realize, the
final gerrymandering model still have some weakness which needs
further improvement.

\subsection{Verification of the Improved Model}
Although we have implemented the improved model to take into
consideration the fairness of voter distribution by party, we cannot
obtain the data on population density favoring respective parities
or on supporting ratio of a specific party. Therefore we can not put
that part of the model into practice. In another word, if we want to
prove the effectiveness and fairness of the model we should try our
best to investigate such data and further correct our simulations
and mathematical analysis.

\subsection{Realization of the Comprehensive Considerations}
Although we have noticed the importance of geographic features when
gerrymandering, due to the complexity of our current model
considering another key factor - avoiding separating administrative
areas, we have not achieved the goal that programs the algorithm
considering the landform modification. Therefore, our next mission
is to realize the comprehensive considerations including geographic
factors.

\newpage

\section{Appendix}
{\bf Applications to Other States}

\begin{figure}[htb]
\begin{center}
  \includegraphics[width= 12 cm]{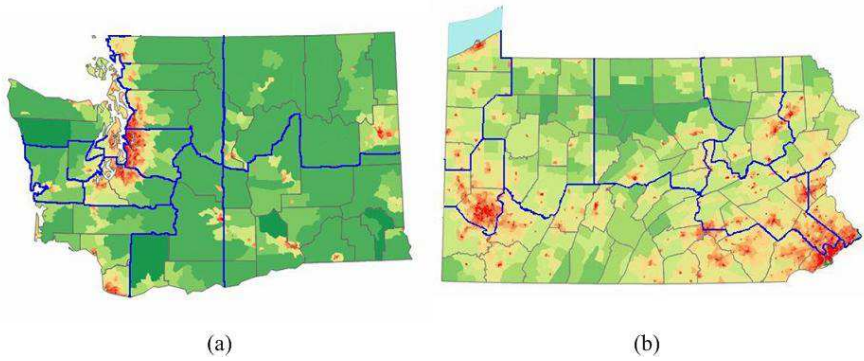}\\
  \caption{\label{wapen} (a)Washington (b)Pennsylvania}
\end{center}
\end{figure}

\begin{figure}[htb]
\begin{center}
  \includegraphics[width= 12 cm]{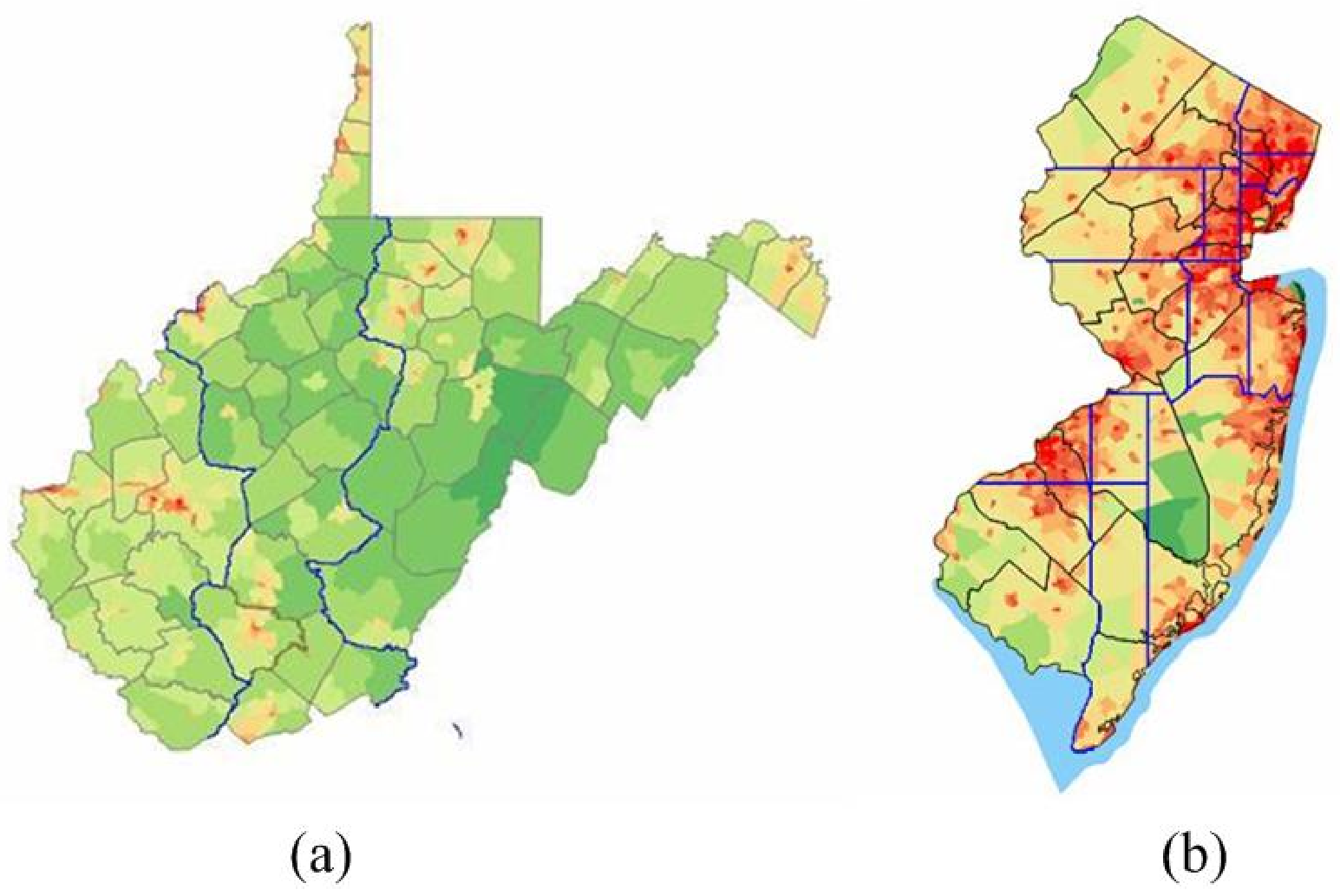}\\
  \caption{\label{wvnj} (a)West Virginia (b)New Jersey}
\end{center}
\end{figure}

\begin{figure}[htb]
\begin{center}
  \includegraphics[width= 12 cm]{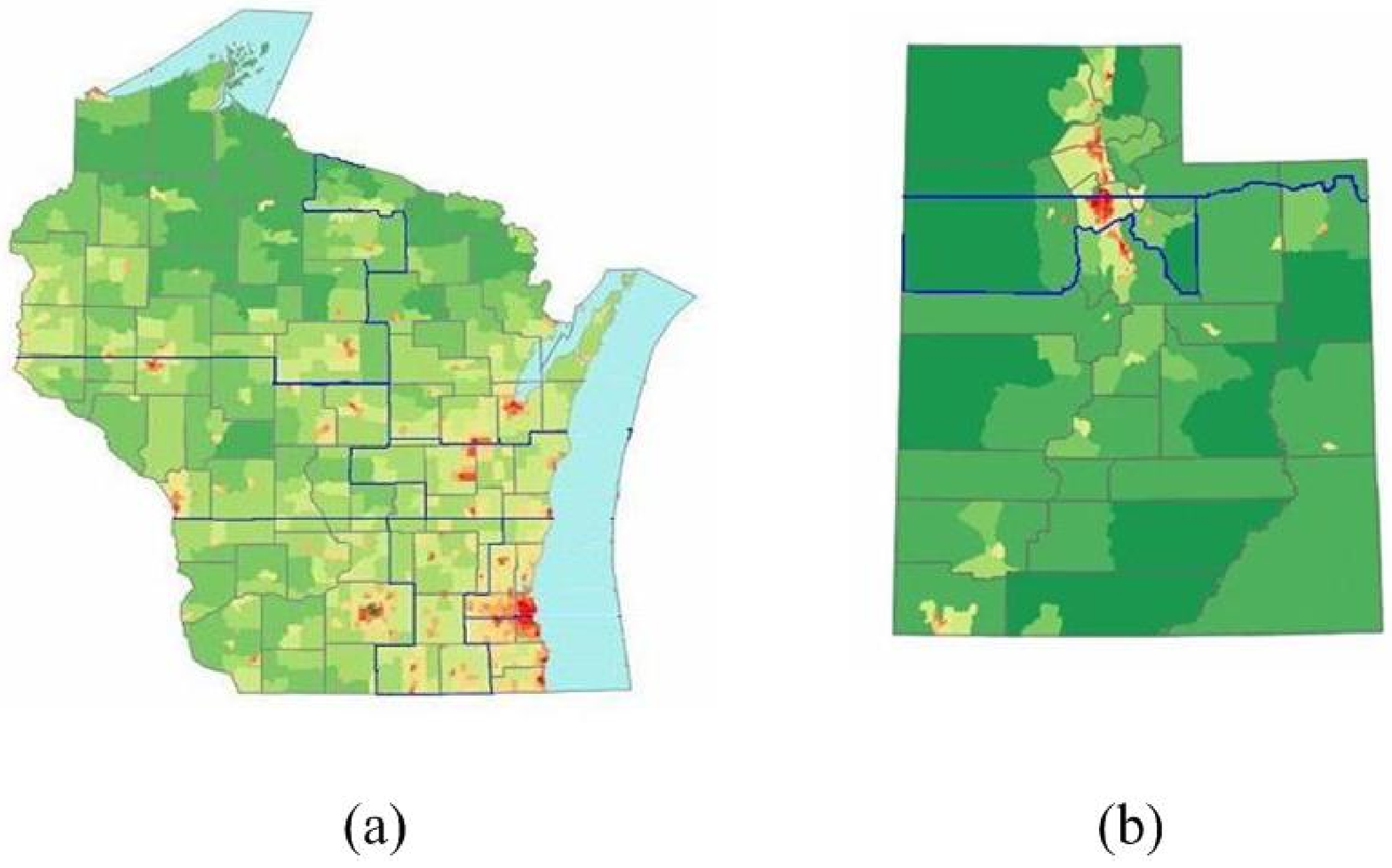}\\
  \caption{\label{wsut} (a)Wisconsin (b)Utah}
\end{center}
\end{figure}

\clearpage
\newpage

\end{document}